\documentclass[aps,twocolumn,showpacs,floatfix]{revtex4}
\usepackage{amsmath}
\usepackage{amsfonts}
\usepackage{amssymb}
\usepackage{epsfig}
\usepackage{graphicx}

\begin{document}
\title{Normal Mode Splitting and Antibunching in Stokes and Anti-Stokes Processes in Cavity Optomechanics: Radiation Pressure Induced Four-Wave Mixing Cavity Optomechanics}
\author{Sumei Huang and G. S. Agarwal}
\affiliation{Department of Physics, Oklahoma State University,
Stillwater, Oklahoma 74078, USA}

\date{\today}

\begin{abstract}
We study Stokes and anti-Stokes processes in cavity optomechanics in
the regime of strong coupling. The Stokes and anti-Stokes signals
exhibit prominently the normal-mode splitting. We report gain for
the Stokes signal. We also report lifetime splitting when the pump
power is less than the critical power for normal-mode splitting. The
nonlinear Stokes processes provide a useful method for studying
strong coupling regime of cavity optomechanics. We also investigate
the correlations between the Stokes and anti-Stokes photons produced
spontaneously by the optomechanical system. At zero temperature, our
nanomechanical system leads to the correlations between the
spontaneously generated photons exhibiting photon antibunching and
those violating the Cauchy-Schwartz inequality.

\end{abstract}
 \maketitle

\renewcommand{\thesection}{\Roman{section}}
\setcounter{section}{0}
\section{Introduction}
\renewcommand{\baselinestretch}{1}\small\normalsize
The nonlinearities in a system can be studied using a number of
optical methods. Among these, Stokes and anti-Stokes processes, and
more generally four-wave-mixing processes are quite common tools
used to understand the nonlinear nature of the system ~\cite{Boyd}.
With this in view we study the stimulated Stokes and anti-Stokes
processes in cavity optomechanics. As is well known, the
nonlinearity in cavity optomechanics arises from the radiation
pressure ~\cite{Meystre,Walther,Fabre,Arcizet,Rokhsari,Mancini} on
the moving mirror of the cavity. Thus, if the cavity is driven by a
pump field of frequency $\omega_{l}$ and a Stokes field of frequency
$\omega_{s}$, then, due to radiation pressure, the output of the
cavity would consist of fields at the applied frequencies
$\omega_{l}$ and $\omega_{s}$ and a generated frequency
$2\omega_{l}-\omega_{s}$. While some previous works
~\cite{Vahala,Braginsky1,Braginsky2} have explored the Stokes and
anti-Stokes processes in the context of parametric oscillation
instability, here we show how such processes can be conveniently
used to study the phenomena of normal-mode splitting
~\cite{Eberly,Agarwal,Raizen,Kimble,Rempe,Klinner,Aspelmeyer,Kippenberg,Sumei}
arising from the strong coupling between the cavity and the
mechanical mirror. Further, the system can act as an amplifier for
the Stokes field. Needless to say, we work in a domain which is
below the instability threshold.

Moreover, very interesting photon correlations between the Stokes
and the anti-Stokes photons have been reported in atomic vapors
under conditions of electromagnetically induced transparency
\cite{Harris}. Here we also discuss the correlations between the
photons created spontaneously by the optomechanical system. The
correlations are found to be nonclassical.

The article is organized as follows. In Sec. II, we introduce the
model, obtain the equation of motion for the oscillator and the
cavity field, and solve it. In Sec. III, we calculate the output
fields and thus obtain nonlinear susceptibilities for Stokes and
anti-Stokes processes. In Sec. IV, we show that the Stokes field is
amplified, and find very prominent normal-mode splittings in the
output fields. Thus, stimulated Stokes and anti-Stokes processes
provide us with a new tool for studying the strong coupling regime
of optomechanics. We find that normal-mode splittings are especially
pronounced in the two quadratures of the output fields. In Sec. V,
we analyze the correlations between the spontaneously generated
photons in the four-wave-mixing processes in the optomechanical
system. We show that such correlations are intrinsically quantum.

\section{Model: Stimulated Generation of Stokes and Anti-Stokes fields}
\begin{figure}[!h]
\begin{center}
\scalebox{0.75}{\includegraphics{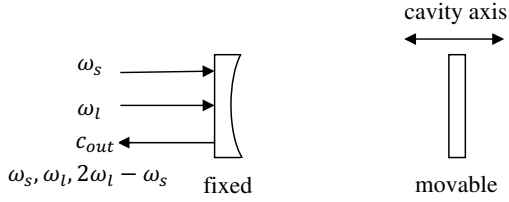}} \caption{\label{Fig1} Sketch
of the studied system. A pump field with frequency $\omega_{l}$ and
a Stokes field with frequency $\omega_{s}$ enter the cavity through
the partially transmitting mirror. The output fields $c_{out}$ have
three components ($\omega_{l},\omega_{s},2\omega_{l}-\omega_{s}$).
No vacuum fields are shown here because we are examining only the
mean response.}
\end{center}
\end{figure}

We consider the system illustrated in Fig.~\ref{Fig1}, in which the
cavity consists of two mirrors separated from each other by a
distance $L$. The front mirror is fixed and partially transmitting;
the end mirror is movable and perfectly reflecting. The cavity is
driven by a pump field and a Stokes field obtained with lasers.
Their frequencies are $\omega_{l}$ and $\omega_{s}$, respectively.
We would assume that the Stokes field is much weaker than the pump
field. A radiation pressure produced by momentum transfer will act
on the movable mirror, which is modeled as a harmonic oscillator
with mass $m$, frequency $\omega_{m}$, and momentum decay rate
$\gamma_{m}$.

Considering a single-mode cavity $\omega_{c}$, the Hamiltonian of
the system in a frame rotating at the pump frequency $\omega_{l}$ is
written as
\begin{equation}\label{1}
\begin{array}{lcl}
\displaystyle
H=\hbar(\omega_{c}-\omega_{l})n_{c}-\hbar\omega_{m}\chi n_{c}
Q+\frac{\hbar\omega_{m}}{4}(Q^2+P^2)\vspace{0.1in}\\\hspace{0.3in}+i\hbar\varepsilon_{l}(c^{\dag}-c)+i\hbar[\varepsilon_{s}
e^{-i(\omega_{s}-\omega_{l})t}c^{\dag}-\varepsilon_{s}^{*}
e^{i(\omega_{s}-\omega_{l})t}c].
\end{array}
\end{equation}
Here $Q$ and $P$ are the dimensionless operators representing the
oscillator's position and momentum, defined by
$Q=q\sqrt{2m\omega_{m}/\hbar}$ and $P=p\sqrt{2/(m\hbar\omega_{m})}$
with $[Q,P]=2i$. In Eq. (\ref{1}), the first term is the energy of
the cavity field, $n_{c}=c^{\dag}c$ is the number of the photons
inside the cavity and $c$ and $c^{\dag}$ are the annihilation and
creation operators, respectively, for the cavity field satisfying
the commutation relation $[c,c^\dag]=1$. The second term describes
the nonlinear coupling of the movable mirror to the cavity field via
radiation pressure, where the dimensionless parameter
$\chi=(1/\omega_{m})(\omega_{c}/L)\sqrt{\hbar/(2m\omega_{m})}$ is
the optomechanical coupling constant between the cavity field and
the movable mirror. The third term corresponds to the energy of the
movable mirror. The last two terms give the interactions of the
cavity field with the pump field and the Stokes field,
$\varepsilon_{l}$ and $\varepsilon_{s}$ are, respectively, the
amplitudes of the pump field and the Stokes field inside the cavity.
They are defined by $\varepsilon_{l}=\sqrt{2\kappa
\wp/(\hbar\omega_{l})}$ and $|\varepsilon_{s}|=\sqrt{2\kappa
\wp_{s}/(\hbar\omega_{s})}$, respectively, where $\wp$ is the pump
power, $\wp_{s}$ is the power of the Stokes field, and $\kappa$ is
the cavity decay rate due to the fixed mirror.

Let $\langle Q \rangle$, $\langle P \rangle$, $\langle c \rangle$,
and $\langle c^{\dag} \rangle$ be the expectation values of the
operators $Q$, $P$, $c$, and $c^{\dag}$, respectively. The time
evolution of these expectation values can be derived by using the
Heisenberg equations of motion and adding the damping terms:
\begin{equation}\label{2}
\begin{array}{lcl}
\langle\dot{Q}\rangle=\omega_{m}\langle P\rangle,\vspace*{.1in}\\
\langle\dot{P}\rangle=2\omega_{m}\chi \langle n_{c}\rangle-\omega_{m}\langle Q\rangle-\gamma_{m}\langle P\rangle,\vspace*{.1in}\\
\langle\dot{c}\rangle=-[\kappa+i(\omega_{c}-\omega_{l}-\omega_{m}\chi
\langle Q\rangle)]\langle c\rangle+\varepsilon_{l}+\varepsilon
_{s}e^{-i(\omega_{s}-\omega_{l})t},\vspace*{.1in}\\
\langle
\dot{c}^{\dag}\rangle=-[\kappa-i(\omega_{c}-\omega_{l}-\omega_{m}\chi
\langle Q\rangle)]\langle
c^{\dag}\rangle+\varepsilon_{l}+\varepsilon
_{s}^{*}e^{i(\omega_{s}-\omega_{l})t}.
\end{array}
\end{equation}
The derivation of Eq. (\ref{2}) uses the well-known mean-field
assumption $\langle Qc\rangle=\langle Q\rangle \langle c\rangle$. As
the field $\varepsilon_{s}$ at the Stokes frequency $\omega_{s}$ is
much weaker than the pump field $\varepsilon_{l}$, we derive the
steady-state solution of Eq. (\ref{2}) to first order in
$\varepsilon_{s}$, that is, we find $t\rightarrow\infty$ limit of
the solutions:
 \begin{equation}\label{3}
 \begin{array}{lcl}
 \left(
  \begin{array}{cccc}
   \langle Q\rangle \\ \langle P\rangle\\  \langle c\rangle \\  \langle c^{\dag}\rangle\\
  \end{array}
\right)=\left(
  \begin{array}{cccc}
   Q_{0}\\ P_{0}\\ c_{0}\\ c_{0}^{*}\\
  \end{array}
\right)+\varepsilon_{s}e^{-i(\omega_s{}-\omega_{l})t}\left(
  \begin{array}{cccc}
    Q_{+} \\ P_{+}\\ c_{+}\\ c_{-}^{*}\\
  \end{array}
\right)\\\hspace{0.7in}+\varepsilon_{s}^{*}e^{i(\omega_s{}-\omega_{l})t}\left(
  \begin{array}{cccc}
   Q_{-} \\ P_{-}\\ c_{-}\\  c_{+}^{*}\\
  \end{array}
\right).
\end{array}
\end{equation}
Thus Eq. (\ref{3}) shows the cavity field $\langle c\rangle
e^{-i\omega_{l}t}$ has three components, oscillating at the input
frequencies $\omega_{l}$ and $\omega_{s}$, and a new anti-Stokes
frequency $2\omega_{l}-\omega_{s}$. By substituting Eq. (\ref{3})
into Eq. (\ref{2}), neglecting those terms containing
$\varepsilon_{s}^2$, $\varepsilon_{s}^{*2}$, and
$|\varepsilon_{s}|^2$  and equating coefficients of terms
proportional to $e^{-i(\omega_s{}-\omega_{l})t}$ and
$e^{i(\omega_s{}-\omega_{l})t}$, respectively, we find

\begin{equation}\label{4}
\begin{array}{lcl}
 Q_{0}= 2\chi|c_{0}|^2\vspace*{.1in},\\
 P_{0}=0,\vspace*{.1in}\\
\displaystyle c_{0}=\frac{\varepsilon_{l}}{\kappa+i\Delta},\vspace*{.1in}\\
\displaystyle
c_{+}=\frac{1}{d(\omega_{s}-\omega_{l})}\{[\kappa-i(\Delta+\omega_{s}-\omega_{l})]\vspace{0.2in}\\\hspace{0.4in}\times[
(\omega_{s}-\omega_{l})^2-\omega_{m}^2+i\gamma_{m}(\omega_{s}-\omega_{l})]\vspace{0.2in}\\\hspace{0.4in}-2i\omega_{m}^3\chi^2|c_{0}|^2\},
\vspace{0.2in}\\
\displaystyle c_{-}=-\frac{2i\omega_{m}^3\chi^2
c_{0}^2}{d^*(\omega_{s}-\omega_{l})}.
\end{array}
\end{equation}
where
\begin{equation}\label{5}
\begin{array}{lcl}
\Delta=\omega_{c}-\omega_{l}-\omega_{m}\chi Q_{0},
\end{array}
\end{equation}
 is the effective detuning, and where
\begin{equation}\label{6}
\begin{array}{lcl}
d(\omega_{s}-\omega_{l})=4\omega_{m}^3\chi^2\Delta
|c_{0}|^2+[(\omega_{s}-\omega_{l}+\omega_m)\vspace{0.2in}\\\hspace{0.7in}\times(\omega_{s}-\omega_{l}-\omega_m)+i\gamma_{m}(\omega_{s}-\omega_{l})
]\vspace{0.2in}\\\hspace{0.7in}\times[\kappa+i(\Delta-\omega_{s}+\omega_{l})][\kappa-i(\Delta+\omega_{s}-\omega_{l})].
\end{array}
\end{equation}
For brevity we do not write explicit expressions for $Q\pm$, $P\pm$,
etc. because we do not need these in the discussion that follows.

\section{The output fields}
To investigate normal-mode splitting of the output fields, we need
to find the expectation value of the output fields. Using
input-output relation \cite{Walls} $\langle
c_{out}\rangle+\varepsilon_{l}/\sqrt{2\kappa}+\varepsilon_{s}e^{-i(\omega_{s}-\omega_{l})t}/\sqrt{2\kappa}=\sqrt{2\kappa}\langle
c\rangle$, we can obtain the expectation value of the output fields
\begin{equation}\label{7}
\begin{array}{lcl}
\langle c_{out}\rangle=\sqrt{2\kappa}[ c_{0}+\varepsilon_{s}
e^{-i(\omega_{s}-\omega_{l})t} c_{+}+\varepsilon_{s}^{*}
e^{i(\omega_{s}-\omega_{l})t}
c_{-}\vspace{0.2in}]\\\hspace{0.5in}-\varepsilon_{l}/\sqrt{2\kappa}-\varepsilon_{s}e^{-i(\omega_{s}-\omega_{l})t}/\sqrt{2\kappa}.
\end{array}
\end{equation}
If we write $\langle c_{out}\rangle$ as
\begin{equation}\label{8}
\langle c_{out}\rangle=c_{l}+\varepsilon_{s}
e^{-i(\omega_{s}-\omega_{l})t} c_{s}+\varepsilon_{s}^{*}
e^{i(\omega_{s}-\omega_{l})t}c_{as},
\end{equation}
where $c_{l}$ is the response at the pump frequency $\omega_{l}$,
$c_{s}$ is the response at the Stokes frequency $\omega_{s}$, and
$c_{as}$ is the response at the four-wave-mixing frequency
$2\omega_{l}-\omega_{s}$ (anti-Stokes frequency).
 Then we have
\begin{equation}\label{9}
\begin{array}{lcl}
\displaystyle \hspace{0.4in}
c_{l}=\frac{\sqrt{2\kappa}\varepsilon_{l}}{\kappa+i\Delta}-\frac{\varepsilon_{l}}{\sqrt{2\kappa}},\vspace{0.2in}\\\hspace{0.4in}
\displaystyle
c_{s}=\frac{\sqrt{2\kappa}}{d(\omega_{s}-\omega_{l})}\{[\kappa-i(\Delta+\omega_{s}-\omega_{l})]\vspace{0.2in}\\\hspace{0.65in}\times[
(\omega_{s}-\omega_{l})^2-\omega_{m}^2+i\gamma_{m}(\omega_{s}-\omega_{l})]\vspace{0.2in}\\\hspace{0.65in}-2i\omega_{m}^3\chi^2|c_{0}|^2\}-\displaystyle
\frac{1}{\sqrt{2\kappa}}, \vspace{0.2in}\\\hspace{0.4in}
\displaystyle c_{as}=-\sqrt{2\kappa}\frac{2i\omega_{m}^3\chi^2
c_{0}^2}{d^*(\omega_{s}-\omega_{l})}.
\end{array}
\end{equation}
In the absence of the interaction between the cavity field and the
movable mirror, one would expect the output fields to contain only
two input components ($\omega_{l}$ and $\omega_{s}$); no
four-wave-mixing component appears.  We can get this result from Eq.
(\ref{9}) by setting $\chi=0$, which  gives
\begin{equation}\label{10}
\begin{array}{lcl}
\displaystyle
\hspace{0.4in}c_{l}=\frac{\sqrt{2\kappa}\varepsilon_{l}}{\kappa+i\Delta}-\frac{\varepsilon_{l}}{\sqrt{2\kappa}},\vspace{0.2in}\\\hspace{0.4in}
\displaystyle
c_{s}=\frac{\sqrt{2\kappa}}{\kappa+i(\Delta-\omega_{s}+\omega_{l})}-\frac{1}{\sqrt{2\kappa}},\vspace{0.2in}\\\hspace{0.4in}
c_{as}=0,
\end{array}
\end{equation}
as expected. However, in the presence of the coupling with the
oscillator ($\chi\ne0$), [from Eq. (\ref{9}), we have $ c_{l}\ne0,
c_{s}\ne0, c_{as}\ne0$], the output fields contain three components.
The generated signal would exhibit resonances whenever
$\omega_{s}=\omega_{l}\pm\omega_{m}$. In addition, one would have
the resonances produced by the cavity
$\omega_{s}=\omega_{l}\pm\Delta$. These resonances are, of course,
expected. The normal-mode splitting would arise as a result of
strong coupling $\chi$ ~\cite{Aspelmeyer, Kippenberg, Sumei}. This
is because the structure of the denominator in Eq. (\ref{9}) depends
on $\chi$. We next present the roots of Eq. (\ref{6}).

We use parameters which have been used in a recent experiment on the
observation of the normal-mode splitting in the fluctuation spectra
\cite{Aspelmeyer}: the wavelength of the laser $\lambda=2\pi
c/\omega_l=1064$ nm, $L=25$ mm, $m=145$ ng,
$\kappa=2\pi\times215\times10^3$ Hz,
$\omega_m=2\pi\times947\times10^3$ Hz, the mechanical quality factor
$Q^{\prime}=\omega_{m}/\gamma_{m}=6700$, $\gamma_{m}=2\pi\times$141
Hz, $\Delta=\omega_{m}$. In this range of parameters, no parametric
instabilities occur.

\begin{figure}[htp]
 \scalebox{0.65}{\includegraphics{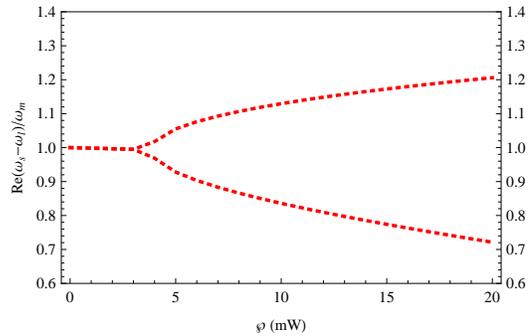}}% Here is how to import EPS art
 \caption{\label{Fig2}(Color online)  The roots of $d(\omega_{s}-\omega_{l})$ in the domain
Re$(\omega_{s}-\omega_{l})>0$ as a function of  the pump power
$\wp$. }
\end{figure}

\begin{figure}[htp]
 \scalebox{0.65}{\includegraphics{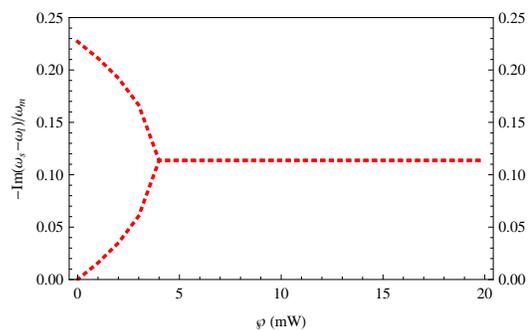}}% Here is how to import EPS art
 \caption{\label{Fig3}(Color online)  The imaginary parts of the roots of $d(\omega_{s}-\omega_{l})$
as a function of the pump power $\wp$.}
\end{figure}

Figure ~\ref{Fig2} shows the dependence of the real parts of the
roots of $d(\omega_{s}-\omega_{l})$ in the domain
Re$(\omega_{s}-\omega_{l})>0$ on the pump power. Figure ~\ref{Fig3}
shows the dependence of the imaginary parts of the roots of
$d(\omega_{s}-\omega_{l})$ on the pump power. For a small value of
the pump power, the real parts of the roots of
$d(\omega_{s}-\omega_{l})$ have two equal values, so there is no
splitting. However, there is lifetime splitting ~\cite{Gupta} as
seen in the Fig.~\ref{Fig3}. If we increase the pump power to a
certain value, the real parts of $d(\omega_{s}-\omega_{l})$ in the
domain Re$(\omega_{s}-\omega_{l})>0$ begin to have two different
values, and the difference between two real parts of the roots of
$d(\omega_{s}-\omega_{l})$ in the domain
Re$(\omega_{s}-\omega_{l})>0$ is increased with increasing pump
power.

\section{Normal-mode splittings in the output fields}

Before examining the normal-mode splitting in, say, the output
anti-Stokes field, we examine Eq. (\ref{9}) in the traditional limit
of nonlinear optics; that is, we find the form of anti-Stokes field
to lowest order in $\chi$,

\begin{widetext}
\begin{equation}\label{11}
\begin{array}{lcl}
\displaystyle c_{as}=-\frac{2\sqrt{2\kappa}i\omega_{m}^3\chi^2
\varepsilon_{l}^2}{(\kappa+i\Delta)^{2}[(\omega_{s}-\omega_{l}+\omega_{m})(\omega_{s}-\omega_{l}-\omega_{m})-i\gamma_{m}(\omega_{s}-\omega_{l})][\kappa-i(\Delta-\omega_{s}+\omega_{l})][\kappa+i(\Delta+\omega_{s}-\omega_{l})]},
\end{array}
\end{equation}
\end{widetext}
which has resonances as discussed after Eq. (\ref{10}) and which is
proportional to the pump power.

\begin{figure}[!h]
\begin{center}
\scalebox{0.7}{\includegraphics{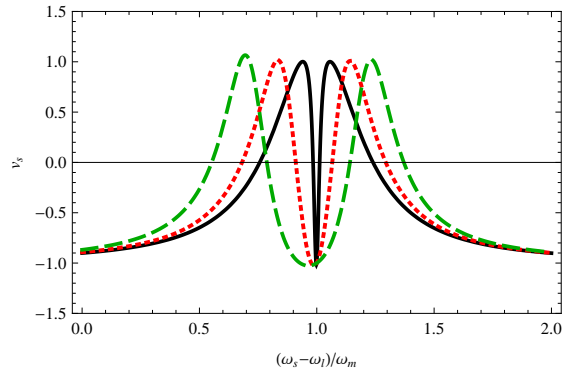}} \caption{\label{Fig4} (Color
online) The normalized quadrature $v_{s}$ plotted as a function of
the normalized frequency $(\omega_{s}-\omega_{l})/\omega_{m}$ for
different pump power. $\wp=1$ mW (solid curve), 6.9 mW (dotted
curve), and 20 mW (dashed curve).}
\end{center}
\end{figure}

\begin{figure}[!h]
\begin{center}
\scalebox{0.7}{\includegraphics{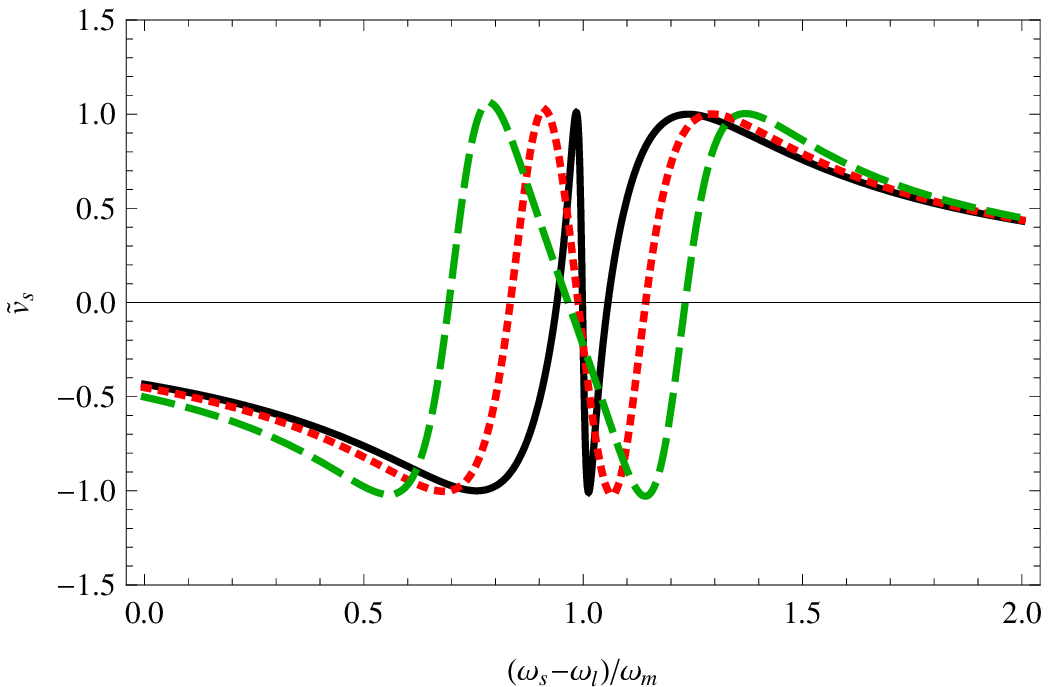}} \caption{\label{Fig5} (Color
online) The normalized quadrature $\tilde{v}_{s}$
 plotted as a function of the normalized frequency
$(\omega_{s}-\omega_{l})/\omega_{m}$ for different pump power.
$\wp=1$ mW (solid curve), 6.9 mW (dotted curve), and 20 mW (dashed
curve).}
\end{center}
\end{figure}

\begin{figure}[!h]
\begin{center}
\scalebox{0.7}{\includegraphics{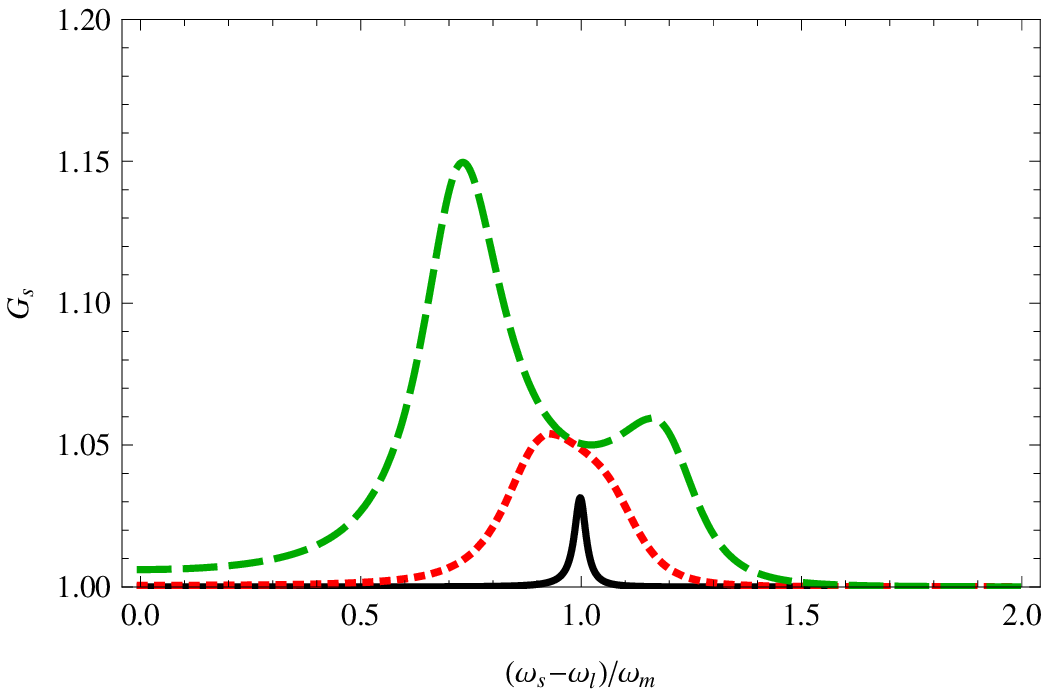}} \caption{\label{Fig6} (Color
online) The normalized output power $G_{s}$ plotted as a function of
the normalized frequency $(\omega_{s}-\omega_{l})/\omega_{m}$ for
different pump power. $\wp=1$ mW (solid curve), 6.9 mW (dotted
curve), and 20 mW (dashed curve).}
\end{center}
\end{figure}

We next discuss the normal-mode splitting in the generated Stokes
and anti-Stokes fields. It is useful to normalize all quantities to
the input Stokes power $\wp_{s}$. For simplicity, we assume
$\varepsilon_{s}$ to be real. For our plots we would give the output
power at the Stokes frequency $\omega_{s}$ in terms of the input
Stokes power
\begin{equation}\label{12}
\begin{array}{lcl}
\displaystyle G_{s}=\frac{\hbar \omega_{s}|\varepsilon_{s}
c_{s}|^2}{\wp_{s}}=|\sqrt{2\kappa} c_{s}|^2,
\end{array}
\end{equation}
and the two quadratures of the output fields at the Stokes frequency
$\omega_{s}$ in terms of the square root of the input Stokes power.
Let us denote these normalized quadratures by $v_{s}$ and
$\tilde{v}_{s}$. These are defined as
$v_{s}=\sqrt{2\kappa}\frac{c_{s}+c^{*}_{s}}{2}$ and
$\tilde{v}_{s}=\sqrt{2\kappa}\frac{c_{s}-c^{*}_{s}}{2i}$. The
quantity $G_{s}$ is the gain of the cavity optomechanical four-wave
mixer. In Figs.~\ref{Fig4}--~\ref{Fig6}, we have plotted $v_{s}$,
$\tilde{v}_{s}$, and $G_{s}$, respectively, versus the normalized
frequency $(\omega_{s}-\omega_{l})/\omega_{m}$ for different pump
powers. The quadrature $v_{s}$ ($\tilde{v}_{s}$) exhibits absorptive
(dispersive) behavior. As is known, there is a phase change on
reflection and that is why the quadrature $v_{s}$ shows absorptive
behavior. The normal-mode splitting or the lifetime splittings are
clearly seen depending on the input pump power in the quadratures
$v_{s}$ and $\tilde{v}_{s}$. The peak positions are in agreement
with Fig.~\ref{Fig2} for the case when the input pump power is such
that normal-mode splitting occurs. The behavior of net gain as a
function of $\omega_{s}$ is different due to the combination of
absorptive and dispersive characteristics of the quadratures $v_{s}$
and $\tilde{v}_{s}$. The gain shows normal-mode splitting for larger
value of the pump power. Moreover, the maximum gain of the Stokes
field is about 1.15. It should be borne in mind that the quadratures
$v_{s}$ and $\tilde{v}_{s}$ can be obtained by homodyne measurement.


\begin{figure}[htp]
\begin{center}
\scalebox{0.7}{\includegraphics{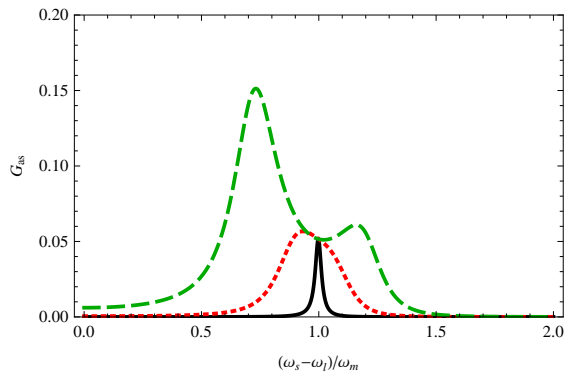}} \caption{\label{Fig7} (Color
online) The normalized output power $G_{as}$ plotted as a function
of the normalized frequency $(\omega_{s}-\omega_{l})/\omega_{m}$ for
different pump power. $\wp=1$ mW (solid curve), 6.9 mW (dotted
curve), and 20 mW (dashed curve). }
\end{center}
\end{figure}

Likewise, the output power at the anti-Stokes frequency
$2\omega_{l}-\omega_{s}$ in terms of the input Stokes power is given
by
\begin{equation}\label{13}
\begin{array}{lcl}
\displaystyle G_{as}=\frac{\hbar
(2\omega_{l}-\omega_{s})|\varepsilon_{s}
c_{as}|^2}{\wp_{s}}=|\sqrt{2\kappa} c_{as}|^2.
\end{array}
\end{equation}
For brevity, we only show in Fig.~\ref{Fig7} the function $G_{as}$
against the normalized frequency
$(\omega_{s}-\omega_{l})/\omega_{m}$ for several values of the pump
power. As can be seen in Fig.~\ref{Fig7}, increasing the pump power
can make the signal of four-wave mixing evolve from one peak to
double peaks. It is also seen that the maximum value of $G_{as}$ is
about 0.15 and the output power at the anti-Stokes frequency
$(2\omega_{l}-\omega_{s})$ is much less than the output power of the
Stokes field. However, for larger pump powers, the maximum gain for
Stokes and anti-Stokes fields are bigger. For example, for 40 mW
pump power, the maximum of $G_{s}$ and $G_{as}$ are about 1.5 and
0.5, respectively.

\section{Spontaneous generation of stokes and anti-stokes
photons: quantum correlations}

So far we have considered stimulated processes. The Stokes and
anti-Stokes fields are also generated spontaneously. In this case we
have to include input vacuum fields. These vacuum fields would be
broad band. Thus the field at frequency $\omega_{s}$ in
Fig.~\ref{Fig1} is to be replaced by a broad band quantum field
$c_{in}$ with zero mean value and with correlations $\langle \delta
c_{in}(t)\delta
c_{in}^{\dag}(t^{\prime})\rangle=\delta(t-t^{\prime})$. The
calculations of the output quantum fields are standard
\cite{Mancini1}. We have used these and introduced the Langevin
force $\xi(t)$ which stems from the coupling of the movable mirror
to the thermal environment having zero mean value with correlations
\cite{Giovannetti}
\begin{equation}\label{14}
\begin{array}{lcl}
\langle \xi(t)\xi(t^{'})\rangle=\displaystyle\frac{1}{\pi}\frac{
\gamma_{m}}{\omega_{m}}\int\omega e^{-i\omega
(t-t^{'})}\left[1+\coth\left(\frac{\hbar
\omega}{2k_{B}T}\right)\right]d\omega,
\end{array}
\end{equation}
where $k_B$ is the Boltzmann constant and $T$ is the temperature of
the environment. The fluctuations of the output fields are obtained
as
\begin{equation}\label{15}
\begin{array}{lcl}
\delta c_{out}(\omega)=V(\omega)\xi(\omega)+E(\omega)\delta
c_{in}(\omega)+F(\omega)\delta c^{\dag}_{in}(-\omega),
\end{array}
\end{equation}
where $\xi(\omega)$, $\delta c_{in}(\omega)$, and $\delta
c_{in}^{\dag}(-\omega)$ are the Fourier transform of the Langevin
force $\xi(t)$ and the input vacuum fields $\delta c_{in}(t)$ and
$\delta c_{in}^{\dag}(t)$, respectively, and where
\begin{equation}\label{16}
\begin{array}{lcl}
V(\omega)=\displaystyle-\frac{\sqrt{2\kappa}\omega_{m}^2\chi}{d(\omega)}i[\kappa-i(\omega+\Delta)]c_{0},\\
E(\omega)=\displaystyle\frac{2\kappa}{d(\omega)}\{-2\omega_{m}^{3}\chi^{2}i|c_{0}|^{2}+(\omega^{2}-\omega_{m}^{2}+i\gamma_{m}\omega)\vspace{.1in}\\
\hspace{.5in}\times[\kappa-i(\omega+\Delta)]\}-1,\vspace{0.1in}\\
F(\omega)=\displaystyle-\frac{4\kappa\omega_{m}^{3}\chi^{2}c_{0}^2}{d(\omega)}i.
\end{array}
\end{equation}
in which
\begin{equation}\label{17}
\begin{array}{lcl}
d(\omega)=4\omega_{m}^3\chi^2\Delta
|c_{0}|^2+(\omega^2-\omega_m^2+i\gamma_{m}\omega
)\vspace{.1in}\\\hspace{.5in}\times[(\kappa-i\omega)^2+\Delta^2].
\end{array}
\end{equation}
In Eq. (\ref{14}), the first term containing $\xi(\omega)$ is the
contribution of the Langevin force acting on the movable mirror,
while the other two terms come from the input vacuum fields. So the
fluctuations of the output fields depend on the Langevin force and
the input vacuum fields. Further, we define time dependent $\delta
c_{out}^{(s)}(t)$ and $\delta c_{out}^{(as)}(t)$, where $\delta
c_{out}^{(s)}(t)$ represents the positive-frequency part of the
fluctuations of the output fields, corresponding to Stokes
component, and
\begin{equation}\label{18}
\delta c_{out}^{(s)}(t)=\frac{1}{2\pi}\int^{\infty}_{0}\delta
c_{out}(\omega)e^{-i\omega t}d\omega,
\end{equation}
whereas $\delta c_{out}^{(as)}(t)$ represents the negative-frequency
part of the fluctuations of the output fields, corresponding to
anti-Stokes component, and
\begin{equation}\label{19} \delta
c_{out}^{(as)}(t)=\frac{1}{2\pi}\int^{0}_{-\infty}\delta
c_{out}(\omega)e^{-i\omega t}d\omega.
\end{equation}

In the context of Stokes and anti-Stokes radiation generated by
single atoms, several authors \cite{Harris, Patnaik, Kimble1,
Scully} found important quantum correlations between the Stokes and
anti-Stokes radiation. Such conclusions were drawn from the
structure of photon-photon correlations. Motivated by these studies
and the fact that we are dealing with a macroscopic system like a
nanomechanical mirror; we examine photon-photon correlations in the
generated radiation.

In the following, like in the work of Kolchin {\it et al.}
\cite{Harris}, we do not differentiate between the Stokes and
anti-Stokes photons. We calculate the coincidence probability
defined by
\begin{widetext}
\begin{equation}\label{20}
\begin{array}{lcl}
g^{(2)}(\tau)=\displaystyle\frac{\langle 0|\delta
c_{out}^{\dag}(t)\delta c_{out}^{\dag}(t+\tau)\delta
c_{out}(t+\tau)\delta c_{out}(t)|0\rangle}{\langle 0|\delta
c_{out}^{\dag}(t)\delta c_{out}(t)|0\rangle\langle 0|\delta
c_{out}^{\dag}(t+\tau)\delta c_{out}(t+\tau)|0\rangle},
\end{array}
\end{equation}
\end{widetext}
in which $\tau$ is a time delay, and
\begin{equation}\label{18}
\delta c_{out}(t)=\frac{1}{2\pi}\int^{+\infty}_{-\infty}\delta
c_{out}(\omega)e^{-i\omega t}d\omega.
\end{equation}

\begin{figure}[htp]
 \scalebox{0.65}{\includegraphics{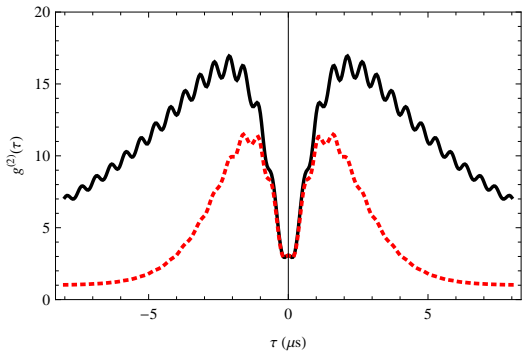}}% Here is how to import EPS art
 \caption{\label{Fig8}(Color online)  The normalized second-order correlation function $g^{(2)}(\tau)$ as a function of the time delay $\tau$($\mu$s) for different pump powers at $T=0$K. $\wp$=1 mW (solid curve), and 4 mW(dotted curve).}
\end{figure}

Now we would evaluate the photon-photon correlations of the output
fields numerically. We choose the pump power $\wp$=1 and 4 mW and
the temperature of the environment $T=0$ K; the other parameters are
the same as those mentioned in Sec. III. The correlation function
$g^{(2)}(\tau)$ between the spontaneously generated photons versus
the time delay $\tau$ for different pump powers at a temperature of
$T=0$K is displayed in Fig.~\ref{Fig8}. We find that $g^{(2)}(\tau)$
is symmetric. It is also seen that $g^{(2)}(\tau)>g^{(2)}(0)$ as
$\tau\neq0$. This demonstrates the presence of photon antibunching,
which is definitely of quantum origin. Further, we note the
Cauchy-Schwartz inequality $g^{(2)}(\tau)\leq g^{(2)}(0)$ is
violated, and the degree of the violation of the Cauchy-Schwartz
inequality becomes smaller with increasing pump power. For pump
power $\wp=1$ mW, the peak value of $g^{(2)}(\tau)$ is about 17, and
$g^{(2)}(0)\approx3$; thus, $g^{(2)}(\tau)/g^{(2)}(0)\approx$5.6.
However, for $\wp=4$ mW, the peak value of $g^{(2)}(\tau)$ is about
11.5, and $g^{(2)}(0)\approx3$, so
$g^{(2)}(\tau)/g^{(2)}(0)\approx$3.8. Therefore, the spontaneously
generated photons from the optomechanical system at $T=0$ K are
correlated nonclassically, and the nonclassical correlation becomes
weaker with increasing pump power. This is reminiscent of the
parametric downconversion process which at low pumping powers
produces significant quantum correlations.

\section{Conclusions}
We have shown that an optomechanical system driven by a pump field
 and a Stokes field can lead to generation of a four-wave-mixing signal. The Stokes field is amplified. We also find that normal-mode splitting occurs in both the generated fields, that is, in both Stokes and anti-Stokes fields. We also report lifetime splitting for pump power less than a
critical power. Further, we have discussed the correlations of the
photons generated from an optomechanical system by spontaneous
processes. We find the correlations between these photons manifest
the antibunching effect, and violate Cauchy-Schwartz inequality.
Further, the violation of the Cauchy-Schwartz inequality becomes
weaker with increasing pump power. Hence, the optomechanical system
can be used to generate pairs of photons with quantum correlations.
Thus the study of both stimulated and spontaneous Stokes and
anti-Stokes signals provides us with a useful technique for studying
the strong coupling regime of cavity optomechanics, as well as
quantum fluctuations at macroscopic level.

We gratefully acknowledge support from the NSF Grant No. PHYS
0653494. We also thank Markus Aspelmeyer for giving us the
experimental data on normal-mode splitting before publication and
for continued correspondence.

\end{document}